# Multispectral palmprint recognition based on three descriptors: LBP, Shift LBP, and Multi Shift LBP with LDA classifier

Salwua Aqreerah
Faculty of Information Technology
University of Benghazi
Benghazi, Libya
salwua.khalifa@uob.edu.ly

Alhaam Alariyibi
Faculty of Information Technology
University of Benghazi
Benghazi, Libya
Alhaam.Alariyibi@uob.edu.ly

Wafa El-Tarhouni
Faculty of Information Technology
University of Benghazi
Libya, Benghazi
wafa.eltarhouni@uob.edu.ly

*Abstract*— Local Binary Patterns (LBP) are extensively used to analyze local texture features of an image. Several new extensions to LBP-based texture descriptors have been proposed, focusing on improving noise robustness by using different coding or thresholding schemes. In this paper we propose three algorithms (LBP), Shift Local Binary Pattern (SLBP), and Multi Shift Local Binary Pattern (MSLBP),to extract features for palmprint images that help to obtain the best unique and characteristic values of an image for identification. The Principal Component Analysis (PCA) algorithm has been applied to reduce the size of the extracted feature matrix in random space and in the matching process; the Linear Discriminant Analysis (LDA) algorithm is used. Several experiments were conducted on the large multispectral database (blue, green, red, and infrared) of the University of Hong Kong. As result, distinguished and high results were obtained where it was proved that, the blue spectrum is superior to all spectra perfectly.

## I. Introduction

Biometric identification systems are considered a fertile area for research to enhance security systems scientific. Palmprint is a relatively new biometric physiological trait that offers stability and uniqueness as part of its traits. The main advantage of a palmprint is the ability to access a large area to extract biometric features. It is worth noting that palm images are often affected by problems when capturing the image, such as noise, contrast, and rotation which making the task of recognition critical. A number of studies have been presented to overcome these problems and provide reliable visibility and fast computations. To date, researchers have developed a number of approaches to extract features and classification with palms. In this research, three algorithms, LBP, SLBP, and MSLBP were used to extract the distinctive and unique features. In order to make the matching process easier, the PCA algorithm was used to reduce the size of the matrix. In the classification stage, the LDA linear discrimination function was applied, which gives the least probability of error. Several experiments have been conducted on the global database PolyU by many researches where in this paper two experiments were conducted. In the first experiment, 3 samples for training and 9 samples for testing were identified. In the second experiment, 6 samples for testing and 6 samples for training were divided; The samples for training were selected randomly.

This paper is organized as follows : Section II is dedicated to presenting some of the research studies concerned with palmprint recognition systems , while section III provides a brief description of the feature extraction, dimensionality reduction, and classification process for the proposed approach. Section IV describes the experimental setup and results, followed by the discussion in the section V. The paper ends with conclusion in Section VI.

## II. Related Works

The human identification system has become a prominent place in the field of research. Where it has led many researchers to focus on proposing different techniques that contribute to further improving the rate of recognition accuracy. Specially that associated with the palm recognition system. In this section, we will provide a brief overview of some of the previous research presented by some researchers on some innovative techniques in the field of human identification using the palm print.

Gumaei et al. [1] suggest an algorithm based on directed gradients (Histogram of Oriented Gradients (HOG)). The extracted features are reduced using Auto-Encoder (AE). The matching process is done using Machine Learning (RELM) Extreme Regularized technology. Experiments were conducted on three palmprint databases, namely MS-PolyU, which holds multispectral palm images, CASIA and Tongji.

An effective classifier for palmprint recognition has been proposed by Rida et al. [2] where it is based on the method of Subspace Random (RSM) technology. The authors used the (2DPCA) Two Dimension Analysis Component Principle algorithm, which is a linear dimension reduction technique, and in the classification process the Neighbors Nearest (NN) algorithm was applied.

In [3], researchers used images with "Bmp" format, being suitable for extracting the main lines and wrinkle lines of the palm of the hand. Then, they are used Contrast Limited Adaptive Histogram Equalization (CLAHE), to improve the contrast of the image. In order to increase the accuracy, a technique is used segmentation, where the "Global Scheme Thresolding" process is applied. After that, the "Gabor Log" technology is used to reduce the unwanted noise from the required information. The experiments of this research were conducted on CASIA database.

Gumaei et al [4], proposed a ROI region segmentation system for images in all spectral bands using Zhang's method, and they used NGist algorithm in the process of image feature extraction, which is an extended version of the Gist algorithm. An automatic encryption algorithm has been used to reduce the properties, and the classification process is done by using the Machine Learning Extreme Regularized (RELM) algorithm. Using the MS-PolyU database for multispectral palm images, several experiments were performed.

Guo et al. [5], a study in which useful information was extracted from a palmprint (non-uniform patterned) pattern to create a bins histogram called multi-hierarchical local binary patterns (HM-LBP). This approach focuses on LBP-based descriptors.

The method for identifying the palm of the hand was proposed by Rida et al. [6], in which the classifier was based on the RSM method, and the 2DPCA algorithm was relied on to build and reduce random subspaces. To extract the features, an algorithm was applied 2DLDA In the classification stage, the NN algorithm was relied on, and the study was based on three general groups, including the POlyU database.

### III. THE PROPOSED APPROACH

In this section, we will present the proposed palmprint recognition system. The structure for palmprint identification should be based on the best spectrum with the highest classification rate. Biometric recognition systems in general, although they differ in techniques, still share five main steps that they must pass through, which are image capture, initial processing, feature extraction, dimensionality reduction and finally classification or matching. The main innovation lies in the introduction of LBP, SLBP and MSLBP descriptors for characteristic extraction followed by PCA to decrease the dimensionality of the extracted palmprint characteristic vectors and the employment of the LDA classifier. Fig. 1 illustrates the steps of the proposed system to identify the palm of the hand. The main components of the proposed palmprint system are described below.

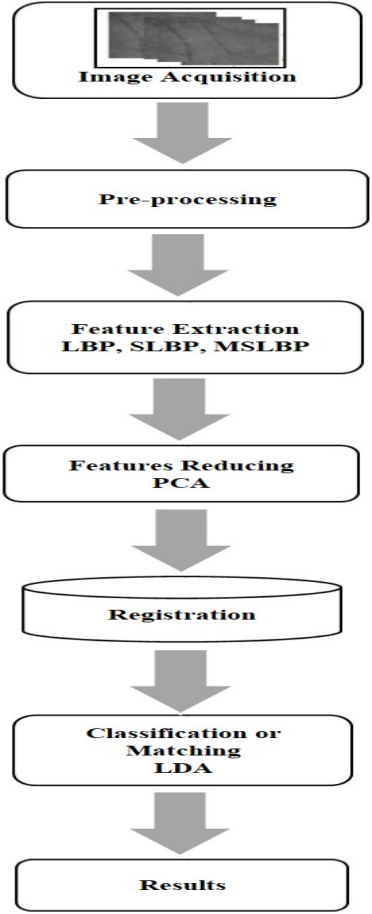

Fig. 1. The process of the proposed system.

### A. Feature Extraction

This stage is one of the important steps for palmprint recognition because it helps to extract some useful features from the palm ROI. The ROI images obtained in the pre-processing stage.

*1) Local Binary Patterns(LBP) :* The original version of the LBP descriptor considers only eight neighbors of a pixel and labels the pixels in the image with the neighborhood threshold of each. Then, it treats the result as a binary number, as originally suggested by Ojala et al. [7]. This demonstrated the great differentiation capacity of this operator to classify textures. Where, the image is divided into (Pixels) cells and the center field and the cells adjacent to the center cell are selected, which in turn will be a square matrix. The adjacent cells are selected by specifying the radius to be dealt with, specifying the number of adjacent cells with a different radius. Each value in the surrounding cell is compared with the value in the center cell. Where the number 1 places for the surrounding cell that has a value greater than or equal to the value carried by the center cell and put the value zero if it is otherwise, as shown in (2). As result, it will be formed a series of binary numbers obtained from these positions and will be converted to a decimal number, and all this may be done based on (1).

$$LBP_{Q,r}(x,y) = \sum_{P=0}^{Q-1} S(g_p - g_c) 2^p \quad (1)$$

$$S(x) = \begin{cases} 1 & if\ x \geq 0 \\ 0 & otherwise \end{cases} \quad (2)$$

Where Q is a set of sample points regularly spaced on a circle of radius r. The value in the central cell c is denoted by $g_c$, and the value in the adjacent cell by $g_p$.

*2) Shift Local Binary Patterns (SLBP):* In the LBP algorithm, one binary code is generated for each pixel in the cell, while in the SLBP algorithm, a number of binary codes is generated for each pixel in the cell and the number of these binary is determined by the variable K, that is determined by (3).

$$k = 2.l + 1 \quad (3)$$

Where

$$k \in [-l, l] \cap z \quad (4)$$

$$SLBP_{Q,r}(x,y,k) = \sum_{p=0}^{Q-1} S(g_p - g_c - k) 2^p \quad (5)$$

For each change in k a new binary code is generated and added to the histogram of patterns [8].

Considering the difference between the LBP and SLBP algorithms lies in the shift value k, the SLBP value is calculated using (5).

And the value obtained from the equation $S(g_p-g_c-k)$ has an impact on choosing the binary number, and S and k can be represented as follows:

$$S(x) = \begin{cases} 1 & if\ x \geq 0 \\ 0 & otherwise \end{cases}\ and\ k \in [-l,l] \cap Z \quad (6)$$

We notice that when the number 3 is assumed as the value of the determinant l, we will get the number 7 as a product of the variable k, which takes the values{ −3, −2, −1, 0, 1, 2, 3 }.

That is, for each pixel, we find that k contributes seven binary codes to the histogram pattern. The final histogram will be divided by k, making the sum of the histogram equal to the number of pixel positions as mentioned in LBP [8, 9].

*3) Multi Shift Local Binary Patterns (MSLBP):* The Local Multi Pattern Binary (MLBP) algorithm has achieved great success on a large scale in distinguishing between different patterns, which has contributed to the interest of many researchers in exploring solutions to problems of contrast (changes) in lighting and rotation that make the task of recognition more complex. Studies have suggested some different algorithms derived from LBP to overcome these difficult problems.

The idea used in MSLBP is to change the radius in the SLBP algorithm, where all non-local information is extracted. Variation in radius depends on the distance of the adjacent cell from the middle cell and allows the formation of a multivariate representation by SLBP binary sequencing. The histogram contains information about the distribution of multivariate features on the entire palm image. Equation (6) was applied at a set of radii (r) as shown in (7) to extract strong patterns from the palm of the hand.

$$H_{Q,r(p)} = \sum_{i=0}^{M-1}\sum_{j=0}^{N-1} g(SLBP_{Q,r}(i,j),p), p \in [1, n+1] \quad (7)$$

Where the histogram is formed as shown in (8) based on the bin patterns obtained from the above equation [10].

$$F_{Q,r} = [H_{Q,1}, H_{Q,2}, \cdots, H_{Q,R}] \quad (8)$$

The resulting multi-scale histogram are used as vectors to be used for the classification stage for the palm of the hand. The algorithm for this method is as follows:-
Step 1: Read the image.
Step 2: Read the radius (r).
Step 3: Extract the palm pattern features by using equations 6, 7, 8.
Step 4: Store palm pattern features in the histogram.
Step 5: If r is less than 8, go to step 2.

*B. Features Dimensionality Reduction*

In palmprint recognition process, there is a problem with the large amounts of data. Dimensional reduction has become a necessity to select the most distinct features for using in the classification stage. The process of reducing statistical dimensional can be used to identify characteristics that are most helpful in the classification process. Principal Component Analysis (PCA) algorithm was used.

*1) Principal Component Analysis (PCA):* This algorithm is a powerful tool for data analysis, and it is used to reduce the dimensions of the data while maintaining the strength of these features to be useful in the classification process. PCA steps can be illustrated as follows:
Step 1: Convert an image to the x, y dimension of the matrix.
Step 2: Calculate the average of each column.
Step 3: Center and standardize the source matrix by subtracting the mean from each column.
Step 4: Calculate the Covariance Matrix.
Step 5: Calculate the eigenvectors and eigenvalues of this covariance matrix.

Step 6: Calculate the projection of the eigenvectors. We chose the projection as a palm feature.

*C. Classification*

Image matching stage, is the process of comparing the test sample and the training set previously stored in the database to identify the person's identity; If a match is found with one of the training samples, the identification process is completed. The proposed approach used the LDA Linear Discriminant Analysis algorithm to perform this stage. It derived the linear discriminant function (LDA) from (Fisher) and it is used when the relationship between the variables is linear [11].

The main concept of LDA is that classes are separated by finding a suitable boundary between them and then classification is performed on the transformed space according to a metric such as Euclidean distance. From a mathematical point of view, the LDA criteria can be satisfied by maximizing the ratio between the intra-class variance determinant and the inter-class variance determinant.

The classification method can be divided into two steps: (1) Calculate the posterior values for each class. And (2) the index of the class to which the test sample belongs in the class with the maximum score is determined by solving the $arg_k max_{g_k}$ [8, 12].

IV. EXPERIMENTAL RESULTS

The POlyU multispectral palm image database [13] was used to test the performance of the proposed system. The database contains 6000 (12*500) images obtained from 500 different palms. So it contains (4*6000) 24000 different palm images. The database contains four types of palm images taken using flexible and infrared light under red, green, blue and NIR illumination. These images were collected in two separate sessions, nine days apart, in each session, 6 images were taken for each of the right and left palms for each of the blue, red, green , and NIR spectral. Two different experiments were applied to evaluation the proposed approach. Fig. 2. Illustrates sample palmprint images of different spectra (red, green, blue and NIR) [15].

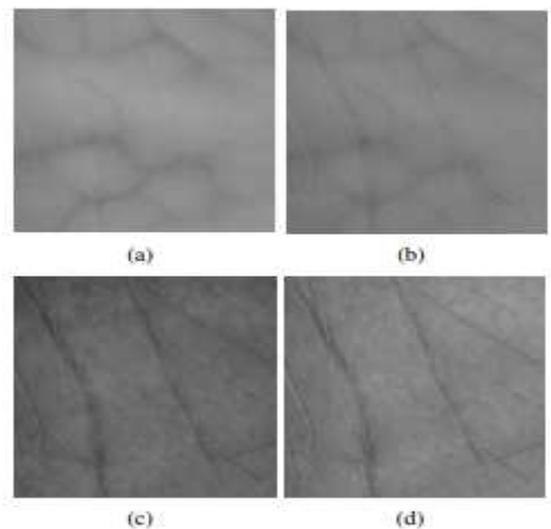

Fig. 2. Sample of ROI palmprint images from PolyU database [15]: (a) NIR, (b) red, (c) green, and (d) blue.

## A. Experiment I

In this experiment, 3 images were selected for each spectrum of palms images (blue, green, red, infrared (NIR)). These images were randomly selected for the training set, and the rest of the images were used as samples for testing set. Where the number of training samples was 1500 images (3×500=1500) while the number of test samples was 4500 (9×500) for each of these spectral (blue, green, red, infrared) to extract the characteristics using the MSLBP algorithm by using different radii (from 1 to 8) and 8 neighbours, the size of the feature matrix extracted from this algorithm was 2048 (8×256), and other algorithms (LBP, SLBP) were also applied in this experiment. The results obtained are presented in Table I, which shows that the MSLBP algorithm, which showed that the MSLBP algorithm had distinguished results with all spectral, as it had the highest rate of identification with the blue spectrum by 99.91% in a time of 24.26 s. As illustrated in Fig. 3, we also obtained good recognition rates with the other spectral, and the less one was with the green spectrum by 99.44% in an average time of 22.17 s. And we also note that it is much higher than the best percentage that we obtained when using the SLBP algorithm, which was with the blue spectrum and is equivalent to 81.36% in a time of 35.47 s.

TABLE I. THE RESULTS OF EXPERIMENT I.

| Methodology | 3 Training and 9 Testing Recognition Rate (%) | | | | | | | |
|---|---|---|---|---|---|---|---|---|
| | Blue | | Green | | Red | | NIR | |
| | Accuracy | Time | Accuracy | Time | Accuracy | Time | Accuracy | Time |
| LBP | 79.44 | 10.67 | 50.42 | 10.32 | 57.44 | 12.86 | 44.69 | 15.12 |
| SLBP | 81.36 | 35.47 | 55.89 | 28.87 | 66.64 | 37.66 | 52.04 | 32.21 |
| MSLBP | 99.91 | 24.26 | 99.44 | 22.17 | 99.87 | 35.11 | 99.69 | 26.42 |

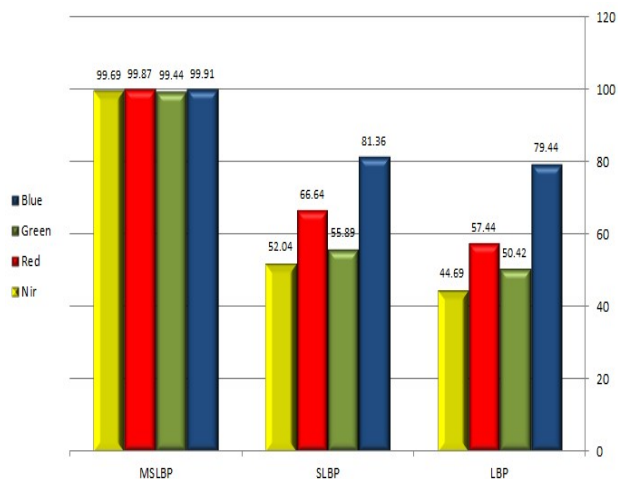

Fig. 3. A comparison of the recognition accuracy of algorithms (LBP, SLBP, MSLBP) obtained in the first experiment using (3 training set and 9 testing set).

## B. Experiment II

In this experiment, 6 images were chosen to be the samples of the training set, and 6 to be the samples for the test set, which was chosen randomly for each spectrum of spectral (blue, green, infrared, red). Here, three algorithms (LBP, SLBP, MSLB) were applied to extract the image features, and after conducting this experiment, the results shown in Table II were obtained, which showed that when applying the LBP algorithm in its simple concept, the results ranged between 64.33% at a time 10.86 seconds with the infrared spectrum and 92.3% in a time of 8.60 seconds with the blue spectrum. As for the SLBP algorithm, these percentages increased to 96,40, 78.83, 67,7,88 μ with the blue, red and green spectral, respectively. However, as shown in Fig. 4, the experiment showed that the MSLBP algorithm had the strongest results, as it reached a good recognition accuracy with the red and infrared spectrum by up to 99.93%, with a difference in time where it was in the red spectrum up to 23.48. While it improved with the infrared spectrum to 20.90, and the obtaining with the blue spectrum a somewhat better recognition accuracy reached 99.96% in a time of 19 seconds, and this algorithm reached its best results with the green spectrum at a rate of 99.9% in 14.81 seconds. By comparing the results of this algorithm with the recognition rate obtained when applying the HOG-SGF+ AE+RELM algorithm[14] and the RSM algorithm[6], the MSLBP algorithm still maintains its strength as the lowest recognition rate still exceeds it with the most powerful access modifier the RSM algorithm, which was with the infrared spectrum by 99.54%, and the percentages were less than that with the rest of the spectral, and the recognition rates were 99.23%, 99.02%, 98.95% with the blue, red and green spectra, respectively.

TABLE II. THE RESULTS OF EXPERIMENT II

| Methodology | 6 Training and 6 Testing | | | | | | | |
|---|---|---|---|---|---|---|---|---|
| | Blue | | Green | | Red | | NIR | |
| | Accuracy | Time | Accuracy | Time | Accuracy | Time | Accuracy | Time |
| LBP | 92.3 | 8.60 | 67.7 | 5.81 | 74.8 | 10.46 | 64.33 | 10.86 |
| SLBP | 96.4 | 22.68 | 78.83 | 21.79 | 88.53 | 28.33 | 78.83 | 26.53 |
| MSLBP | 99.96 | 19.00 | 99.9 | 14.81 | 99.93 | 23.48 | 99.93 | 20.90 |
| HOG-SGF + AE+ RELM[14] | 99.47 | - | 99.40 | - | 99.70 | - | 99.47 | - |
| RSM[6] | 99.23 | - | 98.95 | - | 99.02 | - | 99.54 | - |

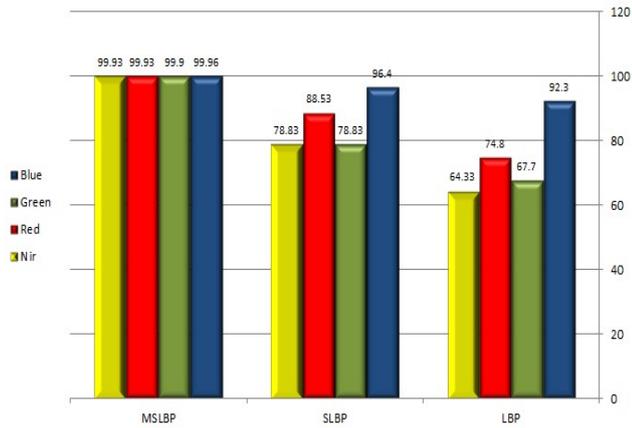

Fig. 4. A comparison of the recognition accuracy of algorithms (L*BP, SLBP, MSLBP*) obtained in the second experiment using (6 training set and 6 testing set).

## V. Discussion

The proposed method successfully captures discriminant information to provide better recognition performance using LDA classification. By looking at the results, it can be seen that the applied MSLBP achieves excellent recognition rates compared to previous studies[6, 14] . In addition, the best result obtained was by applying the MSLBP algorithm with the blue spectrum with a recognition rate of 99.91%, 99.96% in the first and second experiments, respectively, followed by the red spectrum characterized in the two experiments on the recognition rate of the green spectrum as shown in Fig. 5. The important information that must be referred to are the results of the proposed methods for the second experiment (6 Training, 6 Testing) that exceed the results obtained in the first experiment (3 Training, 9 Testing). We can confirm that increasing the number of training samples can improve the performance of the recognition system, also as the Fig. 4 shows the higher the number of features selected, the more accurate the discrimination will be.

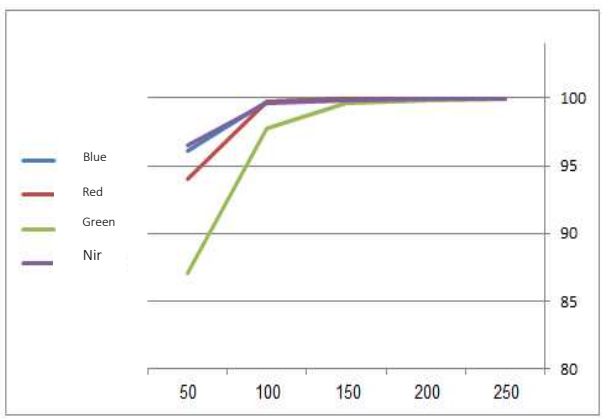

Fig. 5. A digram illustrated the accurancy of the proposed method for the second experiment (6 Training, 6 Testing).

## VI. Conclusion

In this paper, the stage of extracting the characteristics of the palm image is highlighted. Where three algorithms LBP, SLBP, MSLBP were used to extract the distinctive and unique features. Subsequently, all the extracted features are stored in a vector to complete the identification process, and the PCA algorithm is used to reduce the size of the matrix in order to facilitate and succeed in the matching process. In the classification stage, the LDA linear discrimination function was applied, which gives the least probability of error. Several experiments have been conducted on the global database POlyU; Where 3 samples were identified for training and 9 samples for testing the first experiment, while the second experiment was divided into 6 samples for testing and 6 samples for training; the training samples were randomly selected for two experiments. It was concluded that the proposed approach performed better when compared with previous studies. The MSLBP algorithm proved to be highly efficient and accurate in discrimination compared to the LBP and SLBP algorithms.